\documentclass[preprint]{aastex}
\usepackage{color}

\shorttitle{Constraints on feedback processes during the formation of
early-type galaxies}
\shortauthors{Trevisan et al.}

\begin{document}

\title{Constraints on feedback processes during the formation of
early-type galaxies}

\author{M. Trevisan\altaffilmark{1}, I.  Ferreras\altaffilmark{2},
  I.G. de La Rosa\altaffilmark{3,4,5}, F.  La Barbera\altaffilmark{6},
  R.~R. de Carvalho\altaffilmark{7} }
\affil{$^{(1)}$Universidade de S\~ao Paulo/IAG, S\~ao Paulo, Brazil}
\affil{$^{(2)}$MSSL, University College London, Holmbury St Mary,
  Dorking, Surrey RH5 6NT, UK}
\affil{$^{(3)}$Instituto de Astrof\'\i sica de Canarias (IAC), E-38200
  La Laguna, Tenerife, Spain}
\affil{$^{(4)}$Department of Physics and Astronomy, University College
  London, Gower Street, London, WC1E 6BT}
\affil{$^{(5)}$Departamento de Astrofisica, Universidad de La Laguna, E-38205 La Laguna, Tenerife, Spain}
\affil{$^{(6)}$INAF -- Osservatorio Astronomico di Capodimonte,
  Napoli, Italy }
\affil{$^{(7)}$Instituto Nacional de Pesquisas Espaciais/MCT,
  S. J. dos Campos, Brazil}

\email{trevisan@astro.iag.usp.br}

\begin{abstract}
Galaxies are found to obey scaling relations between a number of
observables. These relations follow different trends at the low- and
the high-mass ends. The processes driving the curvature of scaling
relations remain uncertain. In this letter, we focus on the specific
family of early-type galaxies, deriving the star formation histories
of a complete sample of visually classified galaxies from SDSS-DR7
over the redshift range $0.01$~$<$~z~$<$~$0.025$, covering a stellar mass
interval from $10^9$ to $3$~$\times$~$10^{11}$~${\rm M}_{\odot}$. Our sample
features the characteristic ``knee'' in the surface brightness
vs. mass distribution at  M$_\star$~$\sim$~$3$~$\times$~$10^{10}$~M$_\odot$.  We
find a clear difference between the age and metallicity distributions of the
stellar populations above and beyond this knee, which suggests a
sudden transition from a constant, highly efficient mode of star
formation in high-mass galaxies, gradually decreasing towards the
low-mass end of the sample. At fixed mass, our early-type sample is
more efficient in building up the stellar content at early times in
comparison to the general population of galaxies, with half of the
stars already in place by redshift z~$\sim$~2 for all masses. The metallicity-age trend in
low-mass galaxies is not compatible with infall of metal-poor gas,
suggesting instead an outflow-driven relation.
\end{abstract}


\section{Introduction}

The conversion of gas into stars in galaxies and the role of feedback
mechanisms has been one of the key aspects of extragalactic
astrophysics over the past decades. Several baryonic processes
regulate the star formation efficiency within dark matter
haloes. Simulations of galaxy formation and semi-analytical models
have shown that it is only within haloes in a mass range around
M$_{\rm halo}$~$\sim$~M$_{\rm shock}$~$\sim$~$10^{12}$~${\rm M}_{\odot}$ where
baryons can form stars efficiently \citep{Cattaneo.etal:2011,
  Moster.etal:2010, Bouche.etal:2010, Guo.etal:2011}. Above this
limit, gravitational shock heating and AGN feedback suppress the gas
accretion \citep{Dekel.Birnboim:2006, Keres.etal:2009a,
  Keres.etal:2005, Birnboim.Dekel:2003, Cattaneo.etal:2009,
  Cattaneo.etal:2011}. For galaxies within haloes with
masses below $\sim$~$10^{12}$~${\rm M}_{\odot}$, other processes are
usually invoked to explain the star formation
suppression, for instance, reionization of the 
IGM\citep{Mamon.etal:2010, Cattaneo.etal:2011}. The energy liberated
by supernova explosions can eject the gas from haloes with circular
velocity $\lesssim 100$~km~s$^{-1}$, quenching star 
formation \citep{Dekel.Silk:1986}.


The fraction of mass acquired via mergers is also a function of stellar
mass. For example, the semi-analytical models of
\citet{deLucia.etal:2006} show that the number of effective
progenitors of galaxies with stellar masses $\lesssim 10^{11} {\rm
  M}_{\odot}$ is less than two, while this number can be as large as
five for galaxies with ${\rm M}_{\star} \sim 10^{12} {\rm
M}_{\odot}$. According to \citet{Cattaneo.etal:2011}, the dependence of
feedback and merger processes on stellar mass defines three galaxy formation regimes. Stellar mass 
$\sim 10^{11} {\rm M}_{\odot}$ marks the transition between two dominant mechanisms: 
gas accretion (M$_{\star} \lesssim 10^{11} {\rm M}_{\odot}$) and gas-poor mergers (M$_{\star} \gtrsim 10^{11} {\rm M}_{\odot}$). 
A third regime, set immediately below $\sim 10^{11} {\rm M}_{\odot}$, is characterized by the increasing contribution 
of a population built by gas-rich mergers. The contribution of these
mass-dependent processes explain the well known
dichotomy among early-type galaxies \citep[e.g.][]{Kang.etal:2007},
supported by many observations, like e.g. a characteristic mass scale
\citep{Kauffmann.etal:2003}, and a well-defined mass-metallicity
relation \citep[see e.g.][]{Tremonti.etal:2004}.

If the dichotomy originates from the mass-dependent role played by
feedback, gas accretion, gas-rich and gas-poor mergers, we would
expect to find signatures of these processes on the formation history
of galaxies with respect to stellar mass.
In this letter, the star formation history of a sample of galaxies over
a wide range of stellar mass (from $10^9$ to 
$10^{11.5} {\rm M}_{\odot}$) has been examined for the presence of
those signatures.

This letter is organized as follows: in Sect. \ref{Sec_sample}, we
describe the sample; in Sect. \ref{Sec_starlight}, we present a
detailed study of the stellar populations using a spectral fitting
code, which also is able to return the star formation
history. Finally, we summarize and discuss our results in
Sect. \ref{Sec_results}. Throughout the paper, we adopt a cosmology with $H_0
= 70$~km~s$^{-1}$~Mpc$^{-1}$, $\Omega_{\rm m} = 0.3$ and $\Omega_{\Lambda} = 0.7$.

\begin{figure*}
\centering
\begin{tabular}{cc}
   \resizebox{0.35\hsize}{!}{\includegraphics{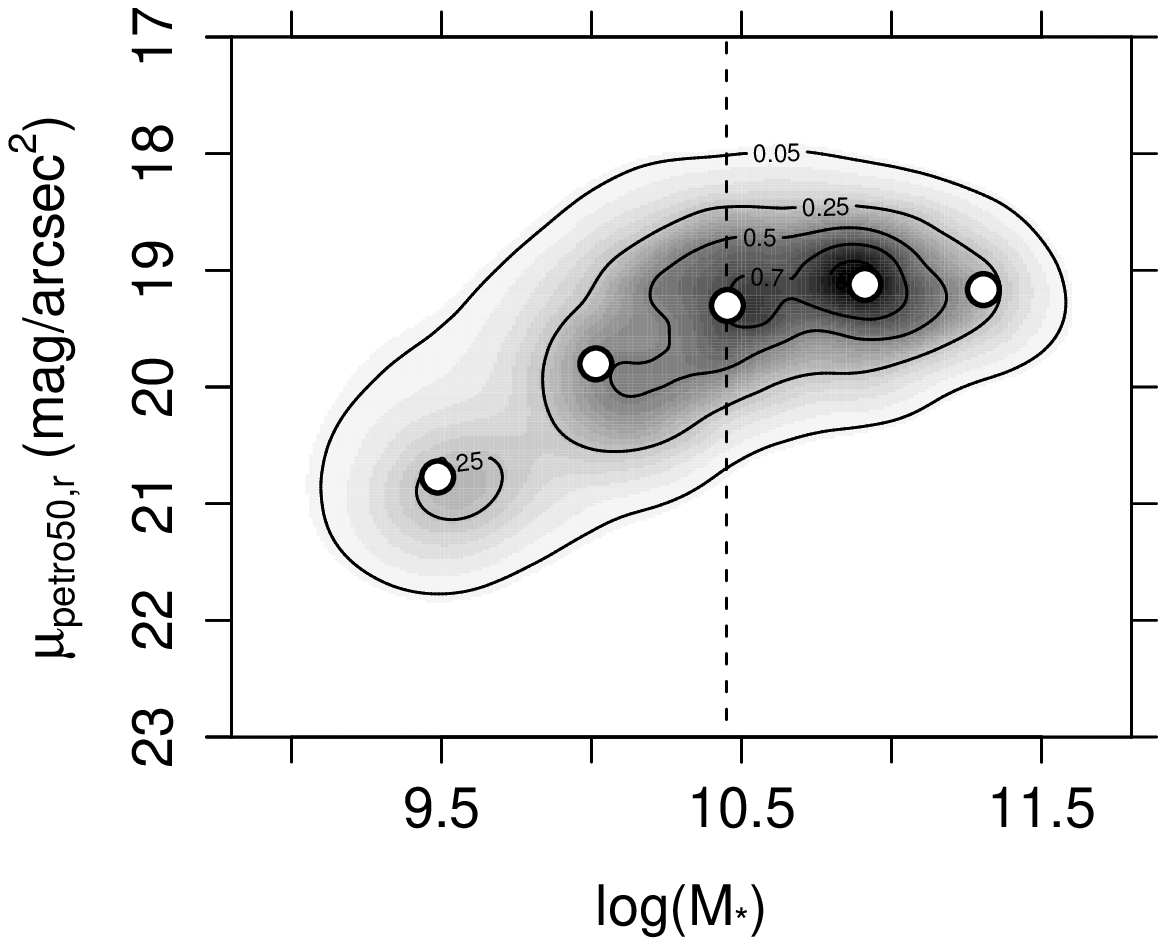}} & 
   \resizebox{0.35\hsize}{!}{\includegraphics{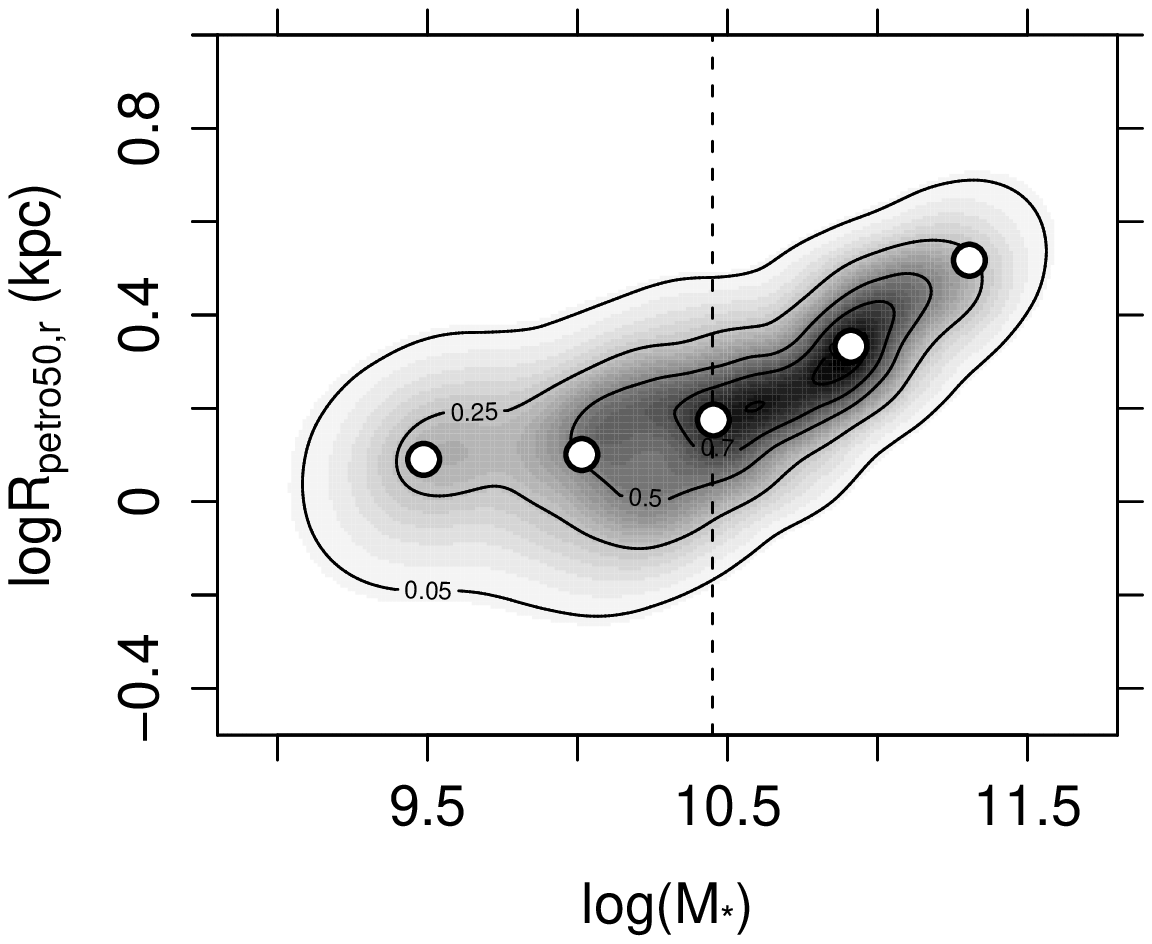}}  \\
   \resizebox{0.35\hsize}{!}{\includegraphics{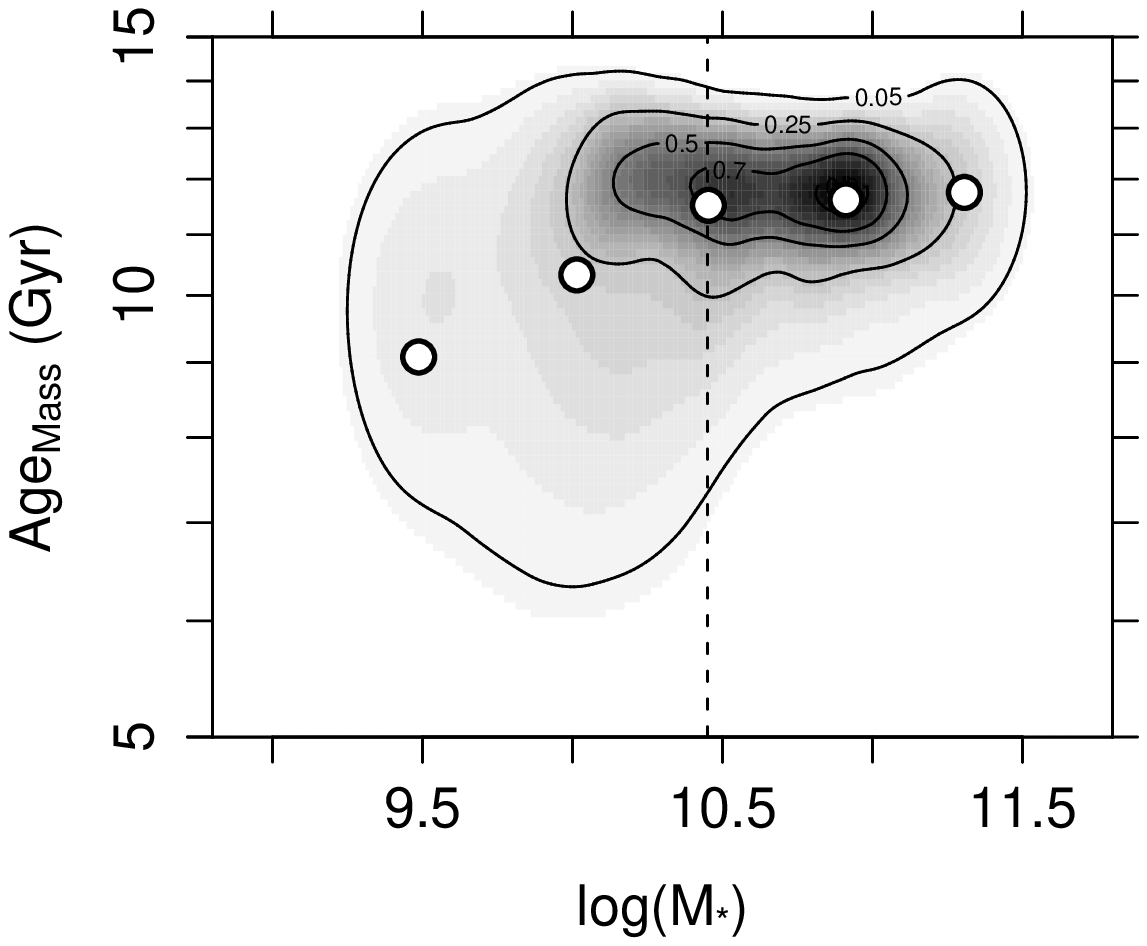}} & 
   \resizebox{0.35\hsize}{!}{\includegraphics{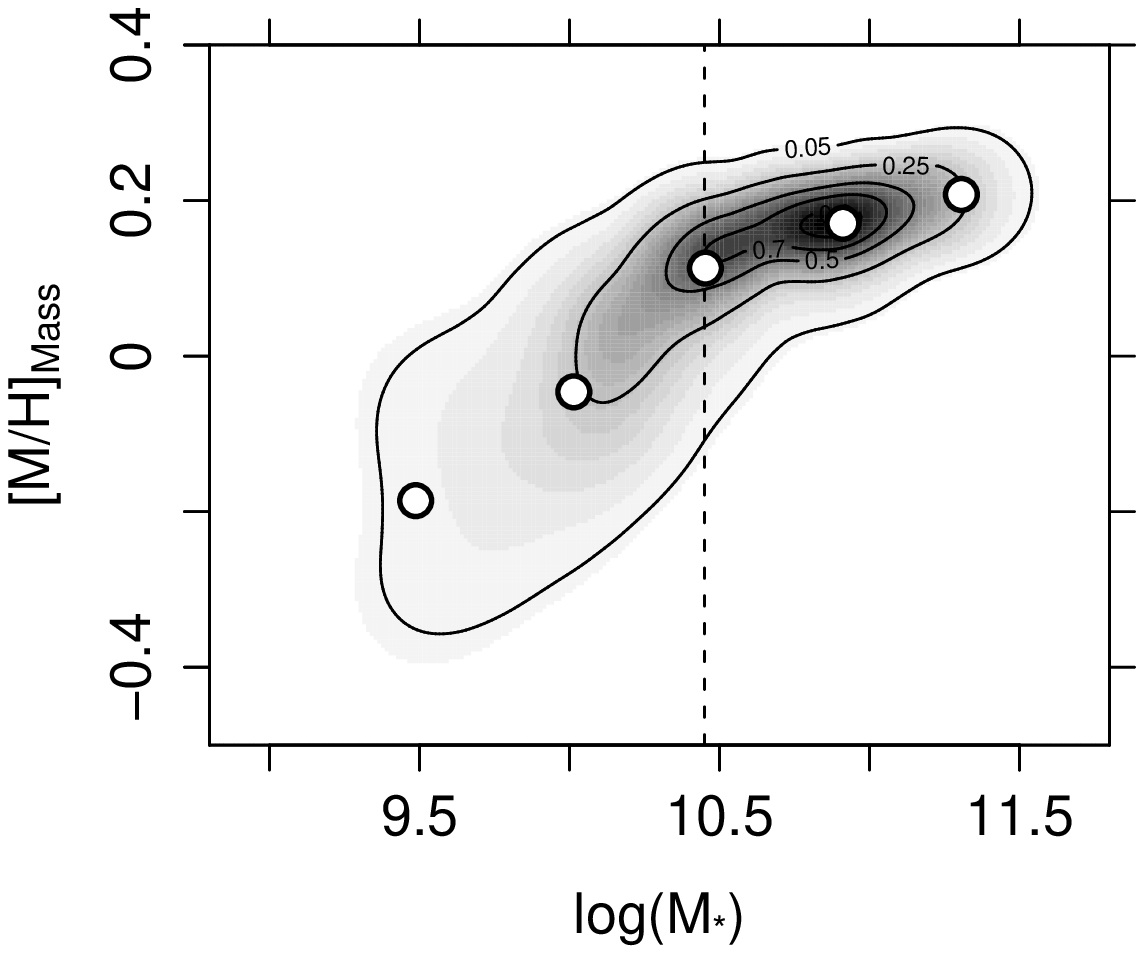}} \\
\end{tabular}
\caption{ {\bf Upper panels:} Scaling relations between stellar mass and
  surface brightness (left) or half-light radius (right). Both quantities are
  measured with respect to the Petrosian radius measured in the $r$
  band.  {\bf Bottom panels:} Mass-weighted average age (left) and
  metallicity (right), as obtained by the {\tt STARLIGHT} spectral fitting
  code (see text for details). The grayscale corresponds to the density of points in
  each graph, with contour lines given at the 90, 70, 50, 25 and 5\%
  levels with respect to the maximum value.  The vertical dashed lines
  indicate M$_{\star} \sim 3 \times 10^{10}$~M$_{\odot}$,
  which corresponds to the stellar mass of objects with
  $\mathcal{M}_{{\rm petro}, r} = -20.2$, i.e. at the knee in the
  scaling relations. Median values obtained within mass bins (see Table \ref{Tab_bins}) are shown in each panel as open dots.}
\label{Fig_scale_relations}
\end{figure*}

\section{Sample description}
\label{Sec_sample}

Our sample of early-type galaxies was retrieved from SDSS-DR7
\citep{Abazajian.etal:2009}, selecting galaxies in the redshift range between
$0.01$ and $0.025$, brighter than $m_{\rm petro, r} < 17.77$, where
$m_{\rm petro, r}$ is the Petrosian magnitude in the r-band. This
limit roughly corresponds to the magnitude at which the SDSS
spectroscopy is complete \citep{Strauss.etal:2002}. The redshift
limits chosen provide a 95\% complete sample between $\mathcal{M}_{\rm
  petro, r} \sim -20$ and $\sim -17.46$, where $\mathcal{M}_{\rm
  petro, r}$ is the k-corrected SDSS Petrosian absolute magnitude in
r-band, obtained with the {\tt kcorrect} code (version 4\_2) of
\citet{Blanton.etal:2003}, choosing as reference the median redshift
of the sample (z$_0$=0.021). See \citet{LaBarbera.etal:2008} for
details on the estimation of the completeness limits.

To select objects from the SDSS database, we define a flag
mask, excluding those objects which are: $i$) blended, i. e.,
with more than one photometric
peak\footnote{\scriptsize Since we are using objects from the primary
  catalog, this selection is equivalent to exclude objects with two or
  more peaks, which were not deblended};
$ii$) too bright (detections of $> 200~\sigma$); $iii$) too
large ($r > 4'$ or a deblend with $r > 1/2$~frame); $iv$) saturated; $v$) located close to the
edge of a frame or $vi$) in a region where the sky measurement failed,
thus the photometry is compromised; objects which $vii$) are part of
the extended wing of a bright star or $viii$) which may be an
electronic ghost of a bright star were also excluded. In addition,
only objects with no spectroscopic warning on (i.e. {\tt zWarning}
attribute set to zero) were selected. This selection returns 10187
objects.

Early-type galaxies obey well-studied scaling relations. Several
morphological classification indicators have been proposed from the
parameters of the SDSS pipeline
\citep[e.g. ][]{Strateva.etal:2001}. However, it is still unclear
whether these indicators can be applied to low mass systems. Previous
work has shown that dwarf elliptical galaxies do not have the same
surface brightness profile as their giant counterparts \citep{Graham.Guzman:2003},
or similar star formation histories \citep{Koleva.etal:2009}, even featuring weak spiral
structures \citep{Lisker:2009}. Hence, we cannot use the standard SDSS
attributes such as {\tt eClass} or {\tt fracDev}. Instead, visual
inspection is the most reliable indicator of an early-type
morphology. We use the classification from the Galaxy Zoo project
\citep{Lintott.etal:2011}, selecting only galaxies classified as
elliptical. We found  10138
($99.7$\%) galaxies from our sample in the Galaxy Zoo database. Among them,
 1359 are classified as ellipticals (spirals and unreliable classifications
are discarded). 
The completeness constraint -- in $\mathcal{M}_{\rm
  petro, r}$ -- results in a final sample of 1328 objects.
A further visual inspection was carried out by the authors
to confirm the morphological classification.


\begin{figure*}[!ht]
\centering
\begin{tabular}{c}
   \resizebox{0.9\hsize}{!}{\includegraphics[angle=-90]{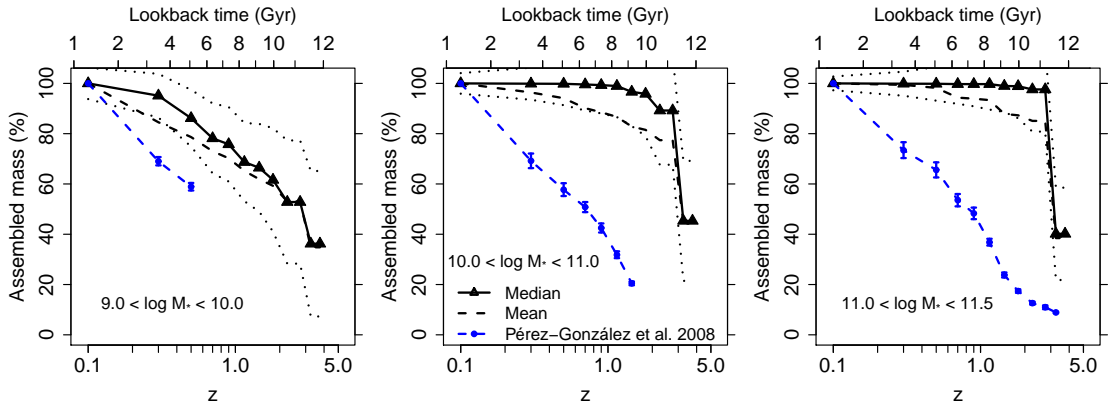}} \\
\end{tabular}
\caption{\label{Fig_t50_t80} Fraction of assembled stellar mass as a
  function of redshift. The panels show the fraction of mass
  normalized by the mass of galaxies at z~$\sim 0.1$. Black triangles and solid (dashed) lines show  median (mean) 
values for the present sample of early-type galaxies, with $1\sigma$
confidence intervals  being marked as dotted lines. Results by
  \citet{PerezGonzalez.etal:2008} are plotted as (blue) solid circles with error bars,
and dashed blue curves.}

\end{figure*}

\section{Stellar content}
\label{Sec_starlight}

Age, metallicity, stellar mass and velocity dispersion were derived
using the spectral fitting code {\tt STARLIGHT}
\citep{CidFernandes.etal:2005}.  Before running the code, the observed
spectra are corrected for foreground extinction and de-redshifted, and
the models are degraded to match the wavelength-dependent resolution
of the spectrum of each galaxy, as described in
\citet{LaBarbera.etal:2010a}.
 
We used SSP models based on the Medium resolution INT Library of
Empirical Spectra \citep[MILES, ][]{SanchezBlazquez.etal:2006}, using the code presented in
\citet{Vazdekis.etal:2010}, using version 9.1 \citep{FalconBarroso.etal:2011}. 
They have a spectral
resolution of $\sim 2.5$ \AA, almost constant with wavelength. A basis
grid of 162 SSPs was selected, covering ages in the range of $0.07 -
12.6$~Gyr, and with [M/H]~=$\{-1.71, -1.31, -0.71, -0.38, 0.00,
+0.20\}$. The models use a \citet{Kroupa:2001} Universal IMF with slope $= 1.30$, and
isochrones by \citet{Girardi.etal:2000}.  The stellar masses --
computed within the fiber aperture -- are extended to the full extent
of the galaxy by computing the difference between fiber and model
magnitudes in the $z$ band.  The stellar mass is then 
$\log($M$_{\star}) = \log({\rm M}_{\star})^{\prime} + 0.4~(m_{{\rm fiber},z} - m_{{\rm
      model},z})$.

We compare the results from {\tt STARLIGHT} with different set-up
parameters and different grids. We find no systematic trends, with
differences typically within $\pm 20$\%. A detailed study of how
results are affected by changes on the set-up parameters of {\tt
  STARLIGHT} and different SSP model assumptions will be given in a
forthcoming paper (Trevisan et al., in prep.). Variations in the SSP
optical colors due to different IMF shapes are very small \citep[see
  e.g. ][]{Vazdekis.etal:2010}.  Since {\tt STARLIGHT} uses all the
spectral information available, different IMFs are not expected to
affect our results.  Hence, if a systematic change of the IMF is
present \citep[as suggested by, e.g.][]{vanDokkum.Conroy:2011} from
low- to high-mass early-type galaxies, the net result would be a
change of the stellar mass that corresponds to the position of the
knee, keeping the derived stellar population properties presented here
unchanged.

\section{Results and discussion}
\label{Sec_results}

In this letter, we study the star formation histories of a sample of
$\sim 1300$ visually-selected elliptical galaxies by means of a
spectral fitting method. For each galaxy we determine the stellar
mass, metallicity, age and star formation history (SFH).
Fig.~\ref{Fig_scale_relations} shows both the photometric scaling
relations (upper panels), along with the scaling of the derived average
ages and metallicities, weighted in mass (bottom panels).  
The plots of surface brightness $\mu_{\rm Petro}$ and Petrosian radius  {\it vs.} stellar
mass provide similar information, considering that $\mu_{\rm Petro}$ is derived from Petrosian 
radius and luminosity, with the latter following the stellar mass of a galaxy. 
We bin the sample into five subsamples in stellar mass, indicated by blue dots in Fig.~\ref{Fig_scale_relations}. 
Table \ref{Tab_bins}
presents the median properties of the stellar populations for each bin.
Regarding the age and metallicity scaling relation (bottom
panels in Fig.~\ref{Fig_scale_relations}), a clear change in the slope of [M/H] {\it vs.} stellar mass
is apparent at  $\sim 3 \times 10^{10} {\rm M}_{\odot}$, equivalent to an absolute magnitude of
$\mathcal{M}_{{\rm dev}, r} \sim -20.2$, below which the metallicity
decreases linearly with mass. This corresponds approximately to the position of the knee seen in
the photometric scale relations.
The age distribution is more complex,
with a homogeneous population of old galaxies for
M$_\star\gtrsim10^{10}$M$_\odot$, and an increased scatter towards
younger ages with decreasing mass.

It is not clear whether the dichotomy in structural properties has the
same origin as the stellar population properties \citep[see
  e.g. ][]{Graham.Guzman:2003, Janz.Lisker:2008, Cote.etal:2008}.  The
fact that the knee in the photometric scaling relations has a
counterpart in the stellar population properties might indicate the
processes regulating the star formation also affect the structural
properties of galaxies.  For example, feedback mechanisms might affect
galaxy sizes, since the gas is pushed out of galaxies by outflows and
might be converted into stars at large radii.  \citep[e.g. ][]{
  Fan.etal:2008, Fan.etal:2010, Damjanov.etal:2009}.
\subsection{Star formation history}
\label{Sec_SFH}

 Besides the averaged age and metallicity, spectral fitting provides a wealth of 
additional information on the star formation histories of individual galaxies.
For a given spectrum, a {\tt STARLIGHT} run returns
the contribution, as a percentage of mass, from each {\it basis} SSP. This distribution
traces directly the star formation history. For each galaxy in the
sample, we determine the ``cumulative'' mass fraction, i.e. the
fraction of stars older than a given age, as a function of
age. Then, we average the cumulative distributions over all galaxies
within each mass bin. The age of the distribution at the 50th and 80th
percentiles in stellar mass is presented in Table
\ref{Tab_bins}. Galaxies with mass $\gtrsim 10^{10} {\rm M}_{\odot}$
form their stars early and over a very short period of time, with 80\%
of their stars being older than $\sim 11$~Gyr. Galaxies in the low
mass bins also have a very old stellar population. However, the time
required to form 80\% of the stellar mass is $\sim 5-6$~Gyr longer
than that required by more massive galaxies. Hence, in early-type
galaxies, {\it downsizing} should be interpreted as a more extended
period of formation in low mass galaxies (instead of an overall later
process of formation). 

We compare our results with \citet{PerezGonzalez.etal:2008}. Their
study is based on a sample of $\sim 28000$ objects of all
morphological types at 0~$<$~z~$<$~4 to constrain the evolution of the
stellar mass content in galaxies as a function of redshift. We rebin
our sample into three subsamples with masses ranging from $\log({\rm
  M}_{\star}) = 9.0-10.0$, $10.0-11.0$, and $11.0-11.5$, which
correspond to the first three bins of
\citet{PerezGonzalez.etal:2008}. Figure \ref{Fig_t50_t80} shows the
cumulative mass fractions as a function of redshift. In all three
bins, early-type galaxies are formed in a much more efficient process,
in contrast to the sample of \citet{PerezGonzalez.etal:2008}, although
the difference is more pronounced in the most massive bin.  The large
difference is due to the fact that their sample includes galaxies of
all morphological types.  In contrast, the analysis of the stellar
populations of visually classified early-type galaxies at
z~$\lesssim$~1, yield a short-lived and early process of star
formation \citep{Ferreras.etal:2009b}, consistent with our findings.
Furthermore, the present sample is susceptible to aperture effects,
since all galaxies are observed through a fiber with fixed angular
diameter.  Table~\ref{Tab_bins} reports the mean aperture for each
mass bin. The aperture $A$ is defined as the ratio between the radius
of the SDSS fiber and the half-light Petrosian radius measured in the
$r$ band, $A = R_{\rm fiber}/R_{{\rm petro50}, r}$.  The mean aperture
varies from $A \sim 0.5$ in the first two bins to $\sim 0.2$ in the
more massive bin.  Assuming that the internal metallicity gradient of
early-type galaxies varies from about $-0.4$~kpc$^{-1}$ at high mass
\citep{LaBarbera.etal:2011} to negligible at lowest mass
\citep[e.g.][]{Koleva.etal:2011}, the above variation of $A$ would
imply a change of $\sim 0.16$ in [M/H], i.e. smaller
than that of  $\sim$~0.4 seen for the range of masses in
Table~\ref{Tab_bins}.   In addition, the typical internal scatter of
  stellar population properties is larger than systematics due to
  aperture effects.  Since age gradients are generally small in
(massive) ETGs, aperture effects are also likely not to drive the
variation of galaxy age with stellar mass. This conclusion is further
supported by our analysis of the waveband dependence of the
Fundamental Plane relation of bright ETGs
\citep{LaBarbera.etal:2010b}, as we found that total (i.e. within an
infinite aperture) metallicity and age do actually increase with
stellar mass.

\begin{table*}[!ht]
\centering
\small
\caption{Stellar population properties as a function of stellar mass.}
 \begin{tabular}{cccccc}
\hline\hline
$\log({\rm M}_{\star})$ interval &  $9.2 - 9.7$ & $9.7 - 10.2$ & $10.2 - 10.7$ & $10.7 - 11.2$ & $11.2 - 11.7$\\
\hline
Number of objects              &  156 & 293 & 410 & 360 & 84 \\
L-weighted Age (Gyr)   &  6.5~$\pm 3.6$ &   8.8~$\pm 3.3$ &   10.3~$\pm 2.3$ &  10.2~$\pm 1.6$ &  9.8~$\pm 1.2$ \\
M-weighted Age (Gyr)   &  9.1~$\pm 2.8$ &  10.3~$\pm 2.4$ &   11.5~$\pm 1.8$ &  11.6~$\pm 1.4$ &  11.8~$\pm 1.2$ \\

50\% of stars older than (Gyr) &  11.2  &  12.4 &  12.4 &  12.4 &  12.4 \\
80\% of stars older than (Gyr) &   5.3  &   8.2 &  11.5 &  11.6 &  11.7 \\
L-weighted [M/H]      &  -0.3~$\pm 0.2$ &  -0.2~$ \pm 0.2$ &  0.0~$ \pm 0.1$ &  0.1~$ \pm 0.1$ &  0.1~$ \pm 0.1$ \\
M-weighted [M/H]      &  -0.2~$\pm 0.2$ &  -0.1~$ \pm 0.1$ &  0.1~$ \pm 0.1$ &  0.2~$ \pm 0.1$ &  0.2~$ \pm 0.1$ \\
$R_{\rm fiber} / R_{{\rm petro50}, r}$ &  0.51~$ \pm 0.19$ &  0.50~$ \pm 0.19$ &  0.42~$ \pm 0.15$ &  0.30~$ \pm 0.09$ &  0.21~$ \pm 0.05$ \\
\hline
\vspace{0.01cm}
 \end{tabular}
\label{Tab_bins}
\end{table*}

\subsection{Constraints of feedback processes}

These results indicate that the processes regulating star formation in
the low- and high-mass regimes leave
different signatures on the SFH. Tab.~\ref{Tab_bins} shows that
galaxies within the three more massive bins have similar ages and
metallicities, with 80\% of their stellar mass formed at approximately
the same redshift. The ages and metallicities are almost
constant from the third to the fifth mass bins. On the other hand, the
low mass bins show a gradual decrease of age and metallicity with
decreasing stellar mass. Galaxies in all bins have roughly half of
their stars formed before redshift z~$\sim 2-3$. For the most massive
galaxies, the additional 30\% of the stars is formed within
$\sim$1~Gyr. Hence, the star formation in these galaxies after a
redshift z~$\sim$~1 can be considered ``residual''. On the other hand,
galaxies with mass ranging from $\log({\rm M}_{\star}) = 9.2$ to $9.7$
have 80\% of their star formed only at z~$\sim 0.5$. 
 Our results indicate that massive objects form faster and low 
mass sytems have a more extended star formation history
than suggested by models \citep[e.g.][]{deLucia.etal:2006}.

The feedback mechanism commonly invoked to explain the regulation of
star formation in low mass systems is supernovae-driven winds
\citep[e.g.][]{Larson:1974,Ferrara.Tolstoy:2000, Keres.etal:2009a}. The chemical
enrichment of low-mas galaxies indicate that the process regulating
the star formation in these galaxies should be strong enough to eject
metals out of the galaxy. However, it is not able to quench the star
formation completely. Alternatively, these systems may be continously
being fuelled by infalling IGM gas at low metallicity.

\begin{figure}[!ht]
\centering
\begin{tabular}{c}
   \resizebox{0.9\hsize}{!}{\includegraphics{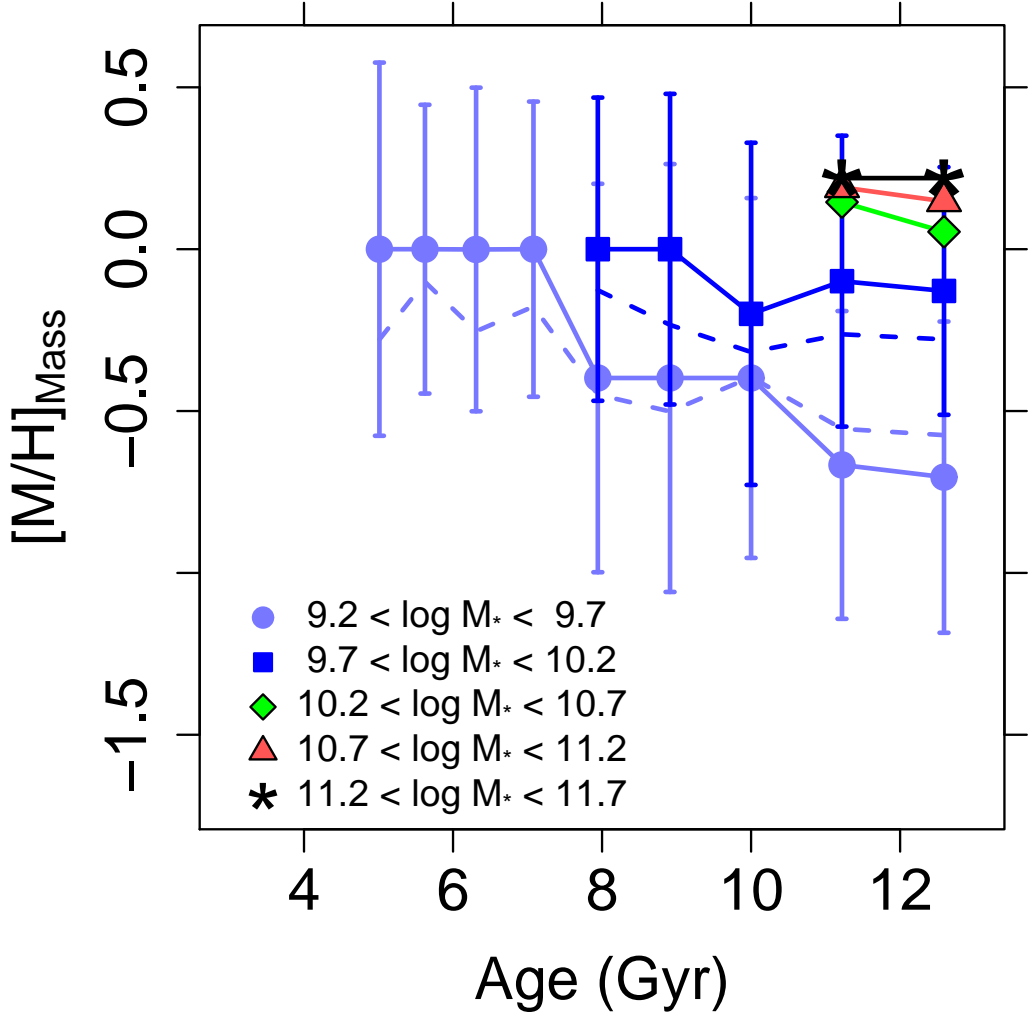}}
   \\
\end{tabular}
\caption{\label{Fig_pop_mets} Mass-weighted metallicity as a function of stellar population age, in bins of stellar mass. 
The metal enrichment is shown until 80\% of the mass is assembled.  For this reason, only two points at $11.2$ and $12.6$~Gyr are shown
for the three most massive bins. This corresponds to the age of the two oldest SSPs used in the spectral fitting.
The solid (dashed) lines indicate the median (mean) values.  Error bars reflect mainly the discreteness of the SSP model grid; 
the metallicities available in the grid are spaced by $\sim$~0.4.}
\end{figure}

Figure \ref{Fig_pop_mets} shows the mass-weighted [M/H] as a function
of stellar population age.  The metal enrichment is shown down to the age when approximately 80\% of the stellar mass is 
formed\footnote{Notice that, for a given mass bin,  the lowest Age in Fig.~3 does not correspond
exactly to the value of $t_{ 80\%}$ in Tab.~1, because of the discreteness of  SSP ages in the STARLIGHT input basis (see Sect.~\ref{Sec_starlight}).}. 
Galaxies in the three most massive bins form their stars very quickly, and 80\% of their
stars are older than 11~Gyr. For this reason, only  two points are shown
for these three bins. The figure shows a consistent trend of chemical
enrichment, whereby the younger populations are more metal rich, a
result that supports the idea that the mass-metallicity trend cannot
be explained by the infall of metal-poor gas, requiring instead a
feedback mechanism that preferentially removes metals from low-mass
galaxies.  Supernova feedback can account for these ``metal-enhanced outflows''.
Since metals are formed in SN events, SN-driven winds are metal-enhanced with respect to the star-forming gas. 
Most of the metals mixed with the hot gas are able to leave the
galaxy, whereas only a small fraction of cool ISM gas is lost 
\citep[see e.g.][]{ Tremonti.etal:2004, MacLow.Ferrara:1999, Ferrara.Tolstoy:2000}.
{ This scenario is compatible with the [M/H] {\it vs.} stellar mass relation shown in 
Fig.~\ref{Fig_scale_relations}. SN-driven feedback is expected to be dominant in low-mass systems,  whereas feedback
processes in massive galaxies leave a sharp truncation in the SFH,
giving rise to different slopes above and below the characteristic mass.
The position of the knee in the photometric scaling relations suggests
that the interplay of these feedback mechanisms leaves an imprint on galaxy sizes and surface brightness.
Aperture effect is not expected to affect these results, 
as galaxies in a given mass range have similar $r_{\rm fiber}/r_{\rm petro}$ ratios. 
 In addition, we verified that the trend is not driven by the age-metallicity degeneracy. We ran STARLIGHT on 
simulated spectra with no chemical-enrichment and similar S/N and star-formation history as galaxies at lowest-mass, 
and we do not find any significant correlation between [M/H]$_{\rm Mass}$ and Age. 


The SFH of high-mass galaxies indicate that their stars were already
formed before z~$\sim 2$. Several studies have shown that high mass
systems are assembled mainly via mergers \citep{deLucia.etal:2006,
  Cattaneo.etal:2011}. However, our results constrain the main epoch
of growth via mergers to z~$\gtrsim$~2, unless the mergers proceed
mainly via gas poor progenitors. Alternatively, these systems can be
already in place at high redshift, as suggested by the weaker number
density evolution with redshift for the most massive galaxies between
z~$\sim 1.5$ and $0$.  \citep[see
  e.g.][]{Conselice.etal:2007,Ferreras.etal:2009,Banerji.etal:2010}.
The cold mode of galaxy formation
\citep{Keres.etal:2005, Guo.etal:2011, Dekel.etal:2009a,
  Dekel.etal:2009b} can account for the required high star formation
efficiency at high redshift. However, it is still not clear how this process can
result in the spheroidal morphologies we see in place at
z~$\sim $~1.

\acknowledgments

\noindent
We thank the referee for very constructive comments that led to
improvements in our manuscript.  MT acknowledges a FAPESP fellowship
no. 2008/50198-3. IGR acknowledges a grant from the Spanish Secretaria
General de Universidades, in the frame of its programme to promote
mobility of Spanish researchers to foreign centers. A full
acknowledgement regarding the use of the Sloan Digital Sky Survey can
be found in this website: {\tt
  http://www.sdss.org/collaboration/credits.html}



\end{document}